\documentclass[%
 reprint,
 amsmath,amssymb,
 aps,superscriptaddress
]{revtex4-2}

\usepackage{graphicx}% Include figure files
\usepackage{dcolumn}% Align table columns on decimal point
\usepackage{bm}% bold math
\usepackage[dvipsnames]{xcolor}

\begin{document}

\preprint{APS/123-QED}

\title{Signature of an Ultrafast Photo-Induced Lifshitz Transition in the Nodal-Line Semimetal ZrSiTe}

\author{Robert J. Kirby}
\affiliation{Department of Chemistry, Princeton University, Princeton, New Jersey 08544, USA}
\author{Lukas Muechler}
\affiliation{Center for Computational Quantum Physics, The Flatiron Institute, New York, New York, 10010, USA}
\author{Sebastian Klemenz}
\affiliation{Department of Chemistry, Princeton University, Princeton, New Jersey 08544, USA}
\author{Caroline Weinberg}
\affiliation{Department of Chemistry, Princeton University, Princeton, New Jersey 08544, USA}
\author{Austin Ferrenti}
\affiliation{Department of Chemistry, Princeton University, Princeton, New Jersey 08544, USA}
\author{Mohamed Oudah}
\affiliation{Department of Chemistry, Princeton University, Princeton, New Jersey 08544, USA}
\affiliation{Quantum Matter Institute, University of British Columbia, Vancouver, British Columbia, V6T 1Z4, Canada}
\author{Daniele Fausti}
\affiliation{Department of Chemistry, Princeton University, Princeton, New Jersey 08544, USA}
\affiliation{Department of Physics, Università degli Studi di Trieste, Trieste I-34127, Italy}
\affiliation{Sincrotrone Trieste S.C.p.A., Basovizza, Trieste I-34012, Italy}
\author{Gregory D. Scholes}
\affiliation{Department of Chemistry, Princeton University, Princeton, New Jersey 08544, USA}
\author{Leslie M. Schoop}
\email{lschoop@princeton.edu}
\affiliation{Department of Chemistry, Princeton University, Princeton, New Jersey 08544, USA}

\date{\today}% It is always \today, today,

\begin{abstract}

Here we report an ultrafast optical spectroscopic study of the nodal-line semimetal ZrSiTe. Our measurements reveal that, converse to other compounds of the family, the sudden injection of electronic excitations results in a strongly coherent response of an $A_{1g}$ phonon mode which dynamically modifies the distance between Zr and Te atoms and Si layers. ``Frozen phonon" DFT calculations, in which band structures are calculated as a function of nuclear position along the phonon mode coordinate, show that large displacements along this mode alter the material's electronic structure significantly, forcing bands to approach and even cross the Fermi energy. The incoherent part of the time domain response reveals that a delayed electronic response at low fluence discontinuously evolves into an instantaneous one for excitation densities larger than $3.43 \times 10^{17}$ cm$^{-3}$. This sudden change of the dissipative channels for electronic excitations is indicative of an ultrafast Lifshitz transition which we tentatively associate to a change in topology of the Fermi surface driven by a symmetry preserving $A_{1g}$ phonon mode.

\end{abstract}

\maketitle

%\tableofcontents

%\section{\label{sec:level1}Introduction}

Light can be a powerful driving force in physical and chemical transformations; photoexcitation above a certain threshold fluence can often induce the same phase transitions at ambient conditions that would normally require, for example, high temperature, such as the metal-to-insulator transition in VO$_2$ \cite{cavalleri2004evidence, kubler2007coherent, wall2012, wegkamp2014instantaneous} or charge-density wave melting in LaTe$_3$ \cite{zong2019evidence}. Importantly, most photo-induced phase transitions (PIPTs) are associated with a structural symmetry breaking. Photoexcitation can also nonadiabatically access regions of a material's potential energy surface that equilibrium stimuli cannot, leading to the possibility of new transformations that would be inaccessible with other external stimuli \cite{ichikawa2011transient, stojchevska2014ultrafast, han2015exploration}. Ultrafast laser experiments give access to out-of-equilibrium PIPTs. Furthermore, ultrafast pump-probe spectroscopic techniques allow one to follow the progress of the PIPT, either directly (as with near-field optical techniques \cite{donges2016}) or indirectly (e.g. through the contribution of certain coherent phonon modes and their fluence dependence \cite{wall2012}). This can provide very powerful insight into the mechanism through which the phase transition occurs, and can help to tailor any potential applications to a certain phase transition and vice versa.

\begin{figure*}
\includegraphics[width=1.0\textwidth]{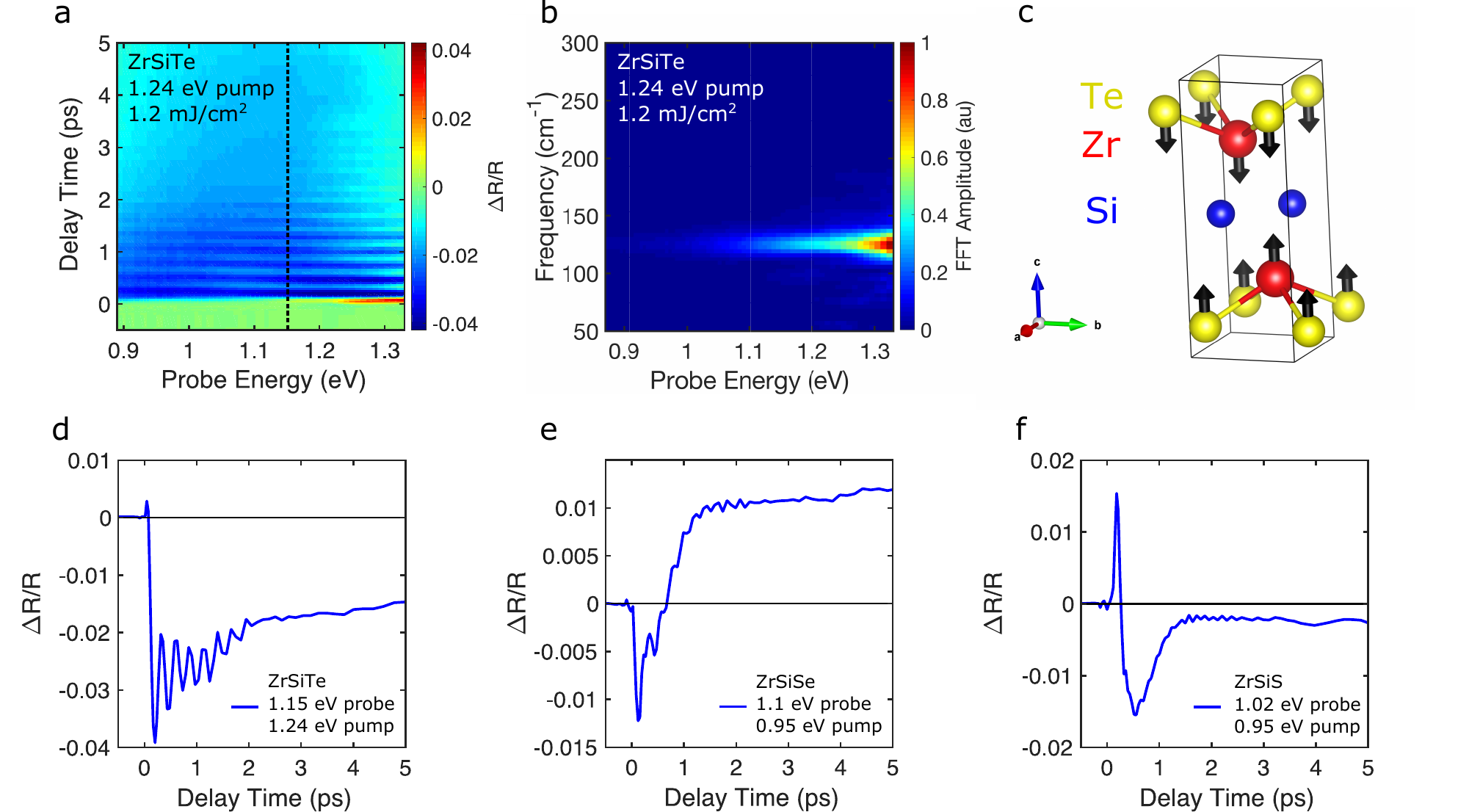}
\caption{\label{fig} \textbf{(a)} Transient response of ZrSiTe pumped at 1.24 eV at a fluence of 1.2 mJ/cm$^2$. The black vertical dashed line indicates the probe energy displayed in \textbf{(d)}. \textbf{(b)} FFT spectrum of the oscillatory response in \textbf{(a)} as a function of probe energy, showing a single frequency at 125 cm$^{-1}$, corresponding to an A$_{1g}$ phonon mode. \textbf{(c)} The crystal structure of ZrSiTe (blue = Si, red = Zr, and yellow = Te) showing one phase of the A$_{1g}$ phonon mode, in which the Zr and Te atoms approach the Si square net from either side; in the other phase of the phonon mode (not shown), the Zr and Te atoms move further away from the Si square net relative to their equilibrium positions. \textbf{(d)} The transient response of ZrSiTe pumped at 1.24 eV, 1.2 mJ/cm$^2$, at a probe energy of 1.15 eV. In the first few periods of the coherent phonon, the amplitude of the oscillatory part is quite large, tens-of-percent of the total magnitude. This is in contrast with the same singular coherent phonon mode observed in ZrSiSe (\textbf{(e)} pumped at 0.95 eV, 1.2 mJ/cm$^2$, probed at 1.1 eV) and ZrSiS (\textbf{(d)} pumped at 0.95 eV, 1.7 mJ/cm$^2$, probed at 1.02 eV). The coherent phonon response in ZrSiTe is significantly stronger than in either ZrSiSe or ZrSiS.  }
\end{figure*}

Recently, indications of a Lifshitz transition \textemdash\ a peculiar electronic phase transition in which the topology of the Fermi surface changes without an accompanying change in crystal symmetry \cite{lifshitz1960anomalies} \textemdash\ have been observed in high-pressure Raman and equilibrium reflectance experiments on the nodal-line semimetal (NLSM) ZrSiTe, supported by density functional theory (DFT) calculations \cite{ebad2019, krottenmuller2020}. In these experiments, pressure was applied along the $c$-axis of the material, forcing the atomic layers of the material closer together. The evidence supporting the electronic phase transitions were discontinuities in the screened plasma frequency (extracted from Kramers-Kronig analysis of the equilibrium reflectance spectra) and the nonlinear response of the frequency of an $A_{1g}$ phonon mode. The DFT calculations indicated that at high pressure, bands cross the Fermi energy along $\Gamma$-X and $\Gamma$-M.

The ZrSi$X$ ($X=$ S, Se, Te) family of materials are the archetypal NLSMs, and have presented themselves as a particularly active area of research in recent years, especially ZrSiTe. Crystallizing in the PbFCl structure type, in the $P4/nmm$ space group, these materials consist of layers of $X$ and Zr on either side of a Si square net, i.e. $X$-Zr-Si-Zr-$X$. Their band structures are characterized by large-bandwidth nodal-line bands near the Fermi energy, which are relatively free from interference from trivial bands. All three ZrSi$X$ compounds also display non-symmorphic symmetry-protected Dirac crossings at the X-point, and although these crossings are hundreds of meV above and below the Fermi energy in ZrSiS and ZrSiSe, they occur at E$_F$ in ZrSiTe \cite{topp2016}. ZrSiTe also has weaker interlayer bonding than the other ZrSi$X$ members \cite{xu2015}, behaving like a quasi-2D material \cite{Hu2016}; it can be exfoliated down to few-layer sheets \cite{yuan2019}, and the monolayer is predicted to be a topological insulator \cite{xu2015}. Additionally, topological drumhead surface states were recently observed for the first time in ZrSiTe \cite{muechler2020}, distinguishing it from ZrSiS where only trivial surface states are experimentally observable\cite{topp2017surface}. Staying within the family, electronic correlations were recently shown to exist in ZrSiSe \cite{shao2020}, and a temperature-induced Lifshitz transition has also been proposed \cite{chen2020} in the same compound.

Other topological semimetals (TSMs) have been shown to exhibit Lifshitz transitions, including ZrTe$_5$ \cite{zhang2017}, VAl$_3$ \cite{liu2020}, NbAs \cite{yang2019}, TaP \cite{caglieris2018}, TaAs \cite{caglieris2018}, WTe$_2$ \cite{wu2015}, NiTe$_2$ \cite{qi2020}, PtTe$_2$ \cite{liu2019}, and T$_{d}$-MoS$_2$ \cite{xu2018}. Most Lifshitz transitions have been observed to occur in equilibrium, induced by a wide range of stimuli including temperature, pressure and strain, and material composition. However, the correlated Weyl semimetal T$_{d}$-MoS$_2$ stands out as being the only material found to undergo a non-equilibrium Lifshitz transition \cite{beaulieu2020}. An ultrafast near-infrared (NIR) laser pulse was used to excite the material, and the subsequent change in the Fermi surface was observed with time- and angle-resolved photoemission spectroscopy (tr-ARPES). Ultrafast spectroscopy has been shown to be such a useful tool in inducing and tracking the evolution of other varieties of electronic and structural phase transitions; it is surprising that there is a dearth of examples of its applicability to Lifshitz transitions.

\begin{figure*}
\includegraphics[width=1.0\textwidth]{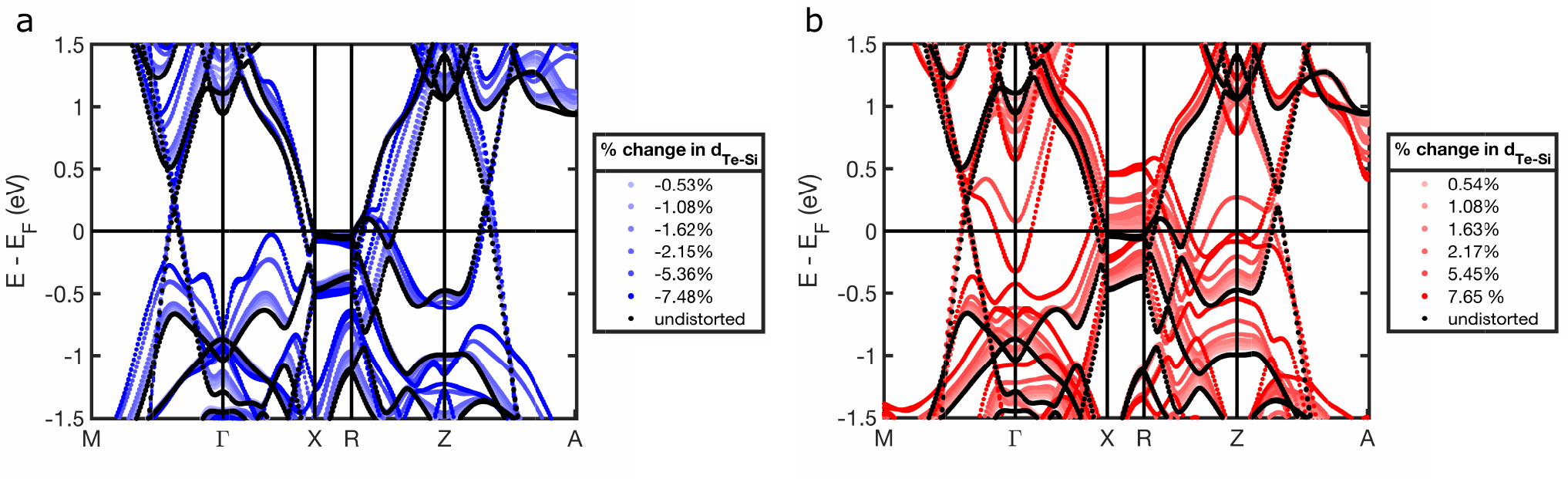}
\caption{\label{fig} Band structures for ZrSiTe for various displacements along the coherent A$_{1g}$ phonon mode. \textbf{(a)} shows the effects of displacements which bring the Zr and Te atoms closer to the Si square net, relative to the equilibrium positions, and \textbf{(b)} shows the effect of the Zr and Te atoms moving further away from the Si square net, relative to their equilibrium positions. The displacement direction in \textbf{(a)} is that displayed in Fig.\ 1 \textbf{(c)}. At high displacements \textemdash\ measured by the change in the Si-Te distance \textemdash\ bands approach and cross E$_F$; ZrSiTe is predicted to undergo Lifshitz transitions along both phases of the coherently-excited A$_{1g}$ phonon mode. }
\end{figure*}

In this letter, we present a combined experimental and computational study of the ultrafast optical response of ZrSiTe, showing that ZrSiTe is only the second topological semimetal to undergo a PIPT. Transient reflectivity shows that a strong relationship exists between the electronic subsystem and the lattice through a particular phonon mode. DFT ``frozen phonon" calculations of the band structure as a function of nuclear position along the phonon mode coordinate show that intense excitation of this phonon mode can have drastic effects on the material's band structure, potentially resulting in a Lifshitz transition. Returning to the transient reflectivity experiments, the signature of an ultrafast PIPT is observed in the fluence dependence of the incoherent part of the transient response, that is, the response of bulk thermalized electrons upon which the oscillatory motion of the coherent phonon is overlaid. While these experiments cannot cast light on the precise nature of the PIPT and the topology of the resulting Fermi surface, the strong excitation and computations suggest that we might be witnessing an ultrafast non-equilibrium Lifshitz transition, as well as a new mechanism by which the transition is mediated by fully-symmetric coherent phonons.

%\section{Experimental Methods}

Transient reflectivity experiments were performed with a commercial pump-probe setup (Ultrafast System Helios, Sarasota, Florida). The 1.55 eV output from a \mbox{1 kHz} regeneratively-amplified Ti:sapphire laser (Coherent Libra, Santa Clara, California) was split with a 50:50 beamsplitter to produce both the pump and probe pulses. Narrowband pump pulses centered at 1.24 and 0.95 eV were produced in an optical parametric amplifier (OPerA Solo, Vilnius, Lithuania). Broadband NIR probe pulses covering 0.8 - 1.35 eV were produced by focusing the \mbox{1.55 eV} light in a sapphire crystal. A $\lambda/2$-waveplate, a polarizer, and an iris control the polarization, intensity, and shape of the beam used to produce the probe pulse. The pulses strike the sample at near-normal incidence, and the reflected probe light is collected by a fiber optic and sent to the detector. All samples were measured in the $ab$ plane, and all measurements were performed at room temperature. Additional details of the sample handling and data processing can be found in the SI.

%\section{Results}

The transient response of ZrSiTe pumped at 1.24 eV with a fluence of 1.2 mJ/cm$^2$ can be seen in \mbox{Fig.\ 1 \textbf{a}}. In this response, large oscillations due to the excitation of coherent phonons are superimposed on top of the bulk incoherent electronic response of the material. The FFT spectrum of the oscillations as a function of probe energy is presented in Fig.\ 1 \textbf{b}, showing that a single phonon mode is being excited; the FFT amplitude increases with increasing probe energy \textemdash\ possibly approaching a resonance \textemdash\ but does not completely disappear at the lower end of the probe window. The coherent phonons dephase after 2 ps.

\begin{figure*}
\includegraphics[width=1.0\textwidth]{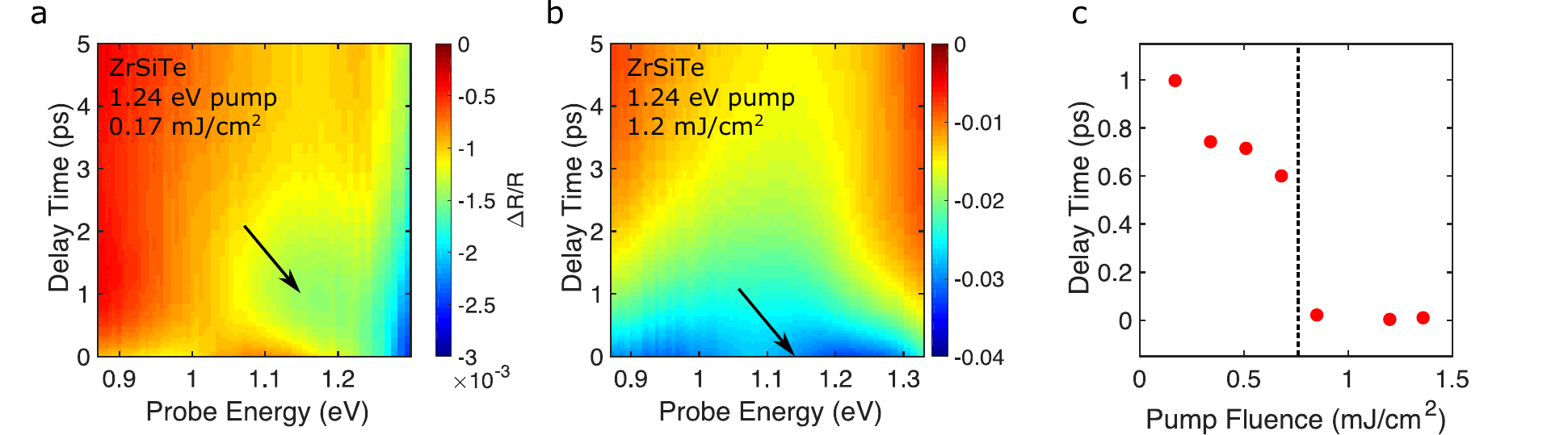}
\caption{\label{fig} The incoherent parts extracted from the transient data for ZrSiTe pumped at 1.24 eV with fluences of 0.17 mJ/cm$^2$ \textbf{(a)} and 1.2 mJ/cm$^2$ \textbf{(b)}. Arrows emphasize the minima in these incoherent parts at 1.15 eV probe energy. \textbf{(c)} illustrates how these incoherent minima fall into two different fluence regimes. Below 0.75 mJ/cm$^2$, the delay time at which the minimum is observed decreases linearly with increasing pump fluence; there is a discontinuity at the threshold fluence 0.75 mJ/cm$^2$, after which the minima are observed at 0 ps delay time, independent of increasing pump fluence. This is the signature of an ultrafast PIPT discussed in the text. }
\end{figure*}

The phonon frequency extracted from the FFT spectrum, 125 cm$^{-1}$, corresponds to an A$_{1g}$ phonon mode also observed in steady-state Raman experiments (Fig.\ S.\ 2); the mode is depicted in Fig.\ 1 \textbf{c}. This mode brings the Zr and Te atoms closer to (Fig.\ 1 \textbf{c}), and farther away from (not shown), the Si square net relative to their equilibrium positions. Almost an order of magnitude of pump fluences were used, ranging from \mbox{0.17 mJ/cm$^2$} to 1.36 mJ/cm$^2$. These experiments were also repeated using a lower energy pump pulse, 0.95 eV. These additional data are presented in Figs.\ S.\ 3, 4, 6, and 7. At both pump energies and at all fluences, the same sole \mbox{125 cm$^{-1}$} Raman mode was observed, with similar relative amplitudes. No additional modes appear above or below a certain fluence, which suggests that ZrSiTe does not undergo a photo-induced change in symmetry within our fluence range. The frequency of the 125 cm$^{-1}$ A$_{1g}$ phonon mode is roughly 15 cm$^{-1}$ higher than that observed in bulk samples in other studies \cite{yuan2019, krottenmuller2020}. This is possibly due to slightly different compositions or carrier concentrations resulting from crystals grown in different labs. However, the frequency is consistent between our transient reflectivity and steady-state Raman measurements (Fig.\ S.\ 2).

The coherent phonon response can be seen more clearly in the transient data at a single probe energy, 1.15 eV, in Fig.\ 1 \textbf{d}. The selfsame A$_{1g}$ coherent phonon mode has also been observed in the transient reflectivity spectra of ZrSiSe and ZrSiS when pumped with 0.95 eV photons (Fig.\ 1 \textbf{e}, \textbf{f}). In these materials, the mode has a frequency of 150 cm$^{-1}$ and 220 cm$^{-1}$, respectively. However, the amplitude of the oscillations in these materials is minuscule compared to those in ZrSiTe, and in experiments in which ZrSiSe and ZrSiS were pumped with \mbox{1.24 eV} photons, no coherent oscillations were observed. Furthermore, these materials only exhibit coherent phonons when pumped at high fluence \textemdash\ 1.2 mJ/cm$^2$ for ZrSiSe and 1.7 mJ/cm$^2$ for ZrSiS \textemdash\ which is not the case for ZrSiTe. Although Te's large polarizability, relative to those of Se or S, is undeniably a contributing factor in the enhanced Raman response compared to the other two materials, it is highly unlikely that this wholly accounts for the significant increase in amplitude. The large amplitude of the coherent phonons relative to the incoherent electronic response \textemdash\ tens of percent in the first period \textemdash\ suggests that this mode is being strongly excited.

A recent study predicts that the geometry of this phonon mode, in which the distance between the Zr and Te atoms and the Si square net is changing, should have a strong impact on the stability of the nodal-line structure \cite{Klemenz2020}. To explore this, ``frozen phonon" DFT calculations were performed using the VASP and phonopy packages~\cite{VASP,phonopy} with the standard pseudopotentials for Zr, Si, and Te. The experimental geometries were taken from Ref.\ \cite{bensch1994structure}. For the self-consistent calculations, the reducible BZ was sampled by a $9\times9\times7$ k-mesh and spin-orbit coupling was included. The one-electron energies were calculated as a function of the $A_{1g}$ normal mode coordinate, assuming the adiabatic Born-Oppenheimer approximation.
% A Wannier interpolation using 64 bands was performed by projecting onto an atomic-orbital basis centered at the atomic positions, consisting of Zr 5$s$,6$s$,5$p$,4$d$,5$d$, Si 3$s$,4$s$,3$p$,4$p$,3$d$ as well as Te 5$s$,6$s$,5$p$,6$p$,5$d$ orbitals.

The results of the calculations are separated by the phase of the phonon motion. The effects of the Te and Zr atoms moving towards the Si square net (compared to their equilibrium positions) on the band structure are shown in Fig.\ 2 \textbf{a}, herein referred to as ``negative distortions", and the effect of the atoms moving further away from the Si square net (again compared to equilibrium), or ``positive distortions", in Fig.\ 2 \textbf{b}. The change in Si-Te distance in the crystal structure due to displacement along the phonon mode is used as a metric for the amount of distortion.

Focusing first on the effects that negative distortions have on the band structure (Fig.\ 2 \textbf{a}), we can see that there are two noticeable changes in the vicinity of the Fermi energy. At high displacement, Te $p$-states increase in energy towards E$_F$ between $\Gamma$ and X, almost touching E$_F$ at the highest amplitude studied. Additionally, the nodal-line bands between R and Z, primarily composed of Si $p$- and Zr $d$-states, shift closer to R and to higher energy even at modest distortions. Both of these band shifts would alter the Fermi surface. Other changes to the band structure occur further away from E$_F$ and are of little consequence to our experimental results.

Moving now to the effects of the positive distortions (Fig.\ 2 \textbf{b}), two striking features stand out. The first can be seen at $\Gamma$, where a band approaches E$_F$ from above, eventually crossing below it at the highest displacement studied. The second is at Z, where bands increase in energy through E$_F$. Other shifts in the energy- and momentum-position of the nodal-line bands can be seen between R and Z, similar to those observed at negative distortions.

These data indicate that ZrSiTe undergoes Lifshitz transitions at the extrema of both phases of the phonon mode observed in the transient data. Due to the strong excitation of this mode, these distortions along the phonon mode are not believed to be unrealistic. The higher negative distortions were chosen to be the Si-$X$, $X=$\ (S, Se), distances in ZrSi$X$ (3.532 \AA\ for ZrSiS and 3.615 \AA\ for ZrSiSe, compared with 3.89105 \AA\ for ZrSiTe), which is a chemically realistic assumption of the range of distortion.

In order to show that the large phonon displacement might drive a Lifshitz transition, we analyze the bulk incoherent electronic response that we mentioned earlier. To extract the incoherent portion from the overall response \textemdash\ comprised of the coherent oscillations and the underlying incoherent part \textemdash\ we fit the transient at each probe energy with a low-order polynomial, which was then subtracted from the overall response leaving just the non-oscillatory part. Fig.\ 3 \textbf{a} and \textbf{b} show these extracted incoherent responses for low (\textbf{a}) and high (\textbf{b}) fluences when pumped at 1.24 eV; the incoherent parts of the other fluences and 0.95 eV experiments can be found in Figs.\ S.\ 5 and 8.

The incoherent part of the low fluence experiment (Fig.\ 3 \textbf{a}) reveals a minimum in the data roughly 1 ps after the pump and probe pulses overlap, at a probe energy of 1.15 eV. However, no such delayed-minimum appears in the high fluence incoherent part (Fig.\ 3 \textbf{b}); here the minimum occurs right at 0 ps. Plotting the delay time at which these minima occur (Fig.\ 3 \textbf{c}), we can see there are two distinct regimes: for fluences below 0.75 mJ/cm$^2$, the delay time minimum decreases linearly with increasing fluence, whereas above 0.75 mJ/cm$^2$ the minimum occurs during excitation by the pump pulse, independent of fluence. The same phenomenon is observed in the \mbox{0.95 eV}-pump experiments, however here the threshold fluence is lower, 0.42 mJ/cm$^2$ (Fig.\ S.\ 6). This behavior, in which there are distinct regimes of a quantity (here the incoherent minimum delay time) separated by a threshold fluence or excitation density, is indicative of a PIPT. The threshold fluences correspond to excitation densities of $3.43 \times 10^{17}$ cm$^{-3}$ (1.24 eV) and $2.71 \times 10^{17}$ cm$^{-3}$ \mbox{(0.95 eV)}. The increase in threshold excitation density with increasing pump energy is almost perfectly mirrored by a decrease in absorbance: 0.0315 (1.24 eV) compared with 0.0417 (0.95 eV), meaning fewer photons need to be absorbed to provide the required energy to complete the phase transition immediately.

%\section{Conclusion}

To conclude, transient reflectivity measurements on the NLSM ZrSiTe show that NIR pulses can strongly excite one particular 125 cm$^{-1}$ A$_{1g}$ phonon mode \textemdash\ coherently \textemdash\ on top of the bulk incoherent electronic response. DFT studies showed that large displacements along this phonon mode adjust the band structure so significantly that bands approach and cross E$_F$, abruptly changing the Fermi surface such that the material undergoes a Lifshitz transition. Furthermore, the fluence dependence of the incoherent part of the transient response has the hallmark of an ultrafast PIPT, with a threshold fluence of 0.75 mJ/cm$^2$ when pumping with 1.24 eV photons and 0.42 mJ/cm$^2$ when pumping with 0.95 eV photons. The large-amplitude coherent phonons give credence to the displacements required to undergo a Lifshitz transition by DFT; however, further studies are necessary to show whether the signature of an ultrafast PIPT observed in the transient data corresponds to this Lifshitz transition or arises from some other ultrafast electronic transition. Moreover, we elucidated this picture within the intuitive Born-Oppenheimer approximation, but in reality the nuclear and electronic degrees of freedom are likely coupled. The possibility of an ultrafast photo-induced Lifshitz transition could make ZrSiTe an ideal candidate material for high-frequency photo-switchable optoelectronic devices, but also presents ZrSiTe as a potential system from which to gain a better understanding of this exciting class of ultrafast PIPT.

\begin{acknowledgments}

This work was supported by NSF through the Princeton Center for Complex Materials, a Materials Research Science and Engineering Center DMR-1420541, by Princeton University through the Princeton Catalysis Initiative, and by the Gordon and Betty Moore Foundation through Grant GBMF9064 to L.M.S. G.D.S. is a CIFAR Fellow in the Bio-Inspired Energy Program. D.F. was supported by the European Commission through the European Research Council (ERC), Project INCEPT, Grant 677488. The authors also acknowledge the use of Princeton's Imaging and Analysis Center, which is partially supported by the Princeton Center for Complex Materials, a National Science Foundation (NSF)-MRSEC program (DMR-1420541). The Flatiron Institute is a division of the Simons Foundation.

\end{acknowledgments}

%\nocite{*}

\providecommand{\noopsort}[1]{}\providecommand{\singleletter}[1]{#1}%

\end{document}